# Security Risks and Modern Cyber Security Technologies for Corporate Networks

Wajeb Gharibi,
College of Computer Science and Information
Systems, Jazan University, Jazan, Saudi Arabia.
gharibi@jazanu.edu.sa

Abdulrahman Mirza,
Center of Excellence in Information Assurance
(CoEIA), King Saud University, KSA.
amirza@ksu.edu.sa

*Abstract*—This article aims to highlight current trends on the market of corporate antivirus solutions. Brief overview of modern security threats that can destroy IT environment is provided as well as a typical structure and features of antivirus suits for corporate users presented on the market. The general requirements for corporate products are determined according to the last report from av-comparatives.org [1]. The detailed analysis of new features is provided based on an overview of products available on the market nowadays. At the end, an enumeration of modern trends in antivirus industry for corporate users completes this article. Finally, the main goal of this article is to stress an attention about new trends suggested by AV vendors in their solutions in order to protect customers against newest security threats.

*Index Terms*—Antivirus technologies, corporate security, corporate network, malicious software, protection, threats, trojan.

## I. INTRODUCTION

MOST companies think of defeating itself against potential security attacks, but only a few of them really imagine a set of security threats that can danger the company. Many of them described in corporate in security standards thus helping the companies to organize IT security defense system. In such context antivirus protection plays the vital part of whole security area. Moreover contemporary antivirus solutions become more advanced and mature. Nowadays they include not only antivirus engine for workstations and an administration console, but many additional features, like antivirus for a mail protection system, a gateway, a database of incidents and enhanced report and logging system. Nonetheless, an implementation of many of such solutions is far from solving all corporate security issues. That is why it is not enough to install only personal antivirus products within a corporate network, but whole corporate suite to cope with all threats at different levels of a network. This will help to construct a corporate secure IT environment.

* This work was partially supported by CoEIA, King Saudi University, KSA.

## II. THE RISK OF MALWARE AND INTERNET THREATS

The main risks for companies in area of information security comprise infections by viruses, trojans, worms, exploits and other malicious code that can reveal the corporate secrets by stealing confidential data and be the reason of serious data leakage. Also phishing and online banking fraud can be a serious problem for IS managers.

Taking in consideration that corporate IT infrastructure mainly consists of domain-joined computers it can be more likely to encounter worms. The main propagation vectors of worms are opened file shares, removable drives, e-mail and IM channels. These are commonly used within companies' networks as a corporate communication and can be a potential threat. According to Microsoft Security Intelligence Report [13], 4 of the top 10 malware families detected on domain-joined computers are worms.

The most popular families are Autorun worms that can spread through removable drives, and network worm Kido/Kido/Conficker/Downadup which was appeared on November 2008 and caused a global world epidemic. The worm has struck more than 10 million computers, using vulnerability in service "Server" (MS08-067).

The worm sent to the remote machine specially crafted RPC-request on TCP port 445 (MICROSOFT_DS|SMB) which caused the buffer overflow by calling wscpy_s() function in NetpwPathCanonicalize() (library netapi32.dll). The given malicious program applied a wide spectrum of methods to hide the presence in the system: files view settings in Explorer, disabling the services, responsible for system security. It was used several ways of distribution: the admin shared folders, removable devices, downloading the updates from websites, domain addresses of which were generated by special algorithm. As a result it has received a wide proliferation all over the Internet. The detailed description of the worm you can find in malware encyclopedia [14].

## III. SECURITY RISKS IN HARDWARE

### A. Overview

Recently Dell Company, the leading computer system manufacturer, announced that in its servers' line PowerEdge malicious program has been found embedded in a flash memory of a motherboard [15].

Thus, the computer industry has been faced with the threat of computers' infection with malicious software, but at the level of firmware. The topic of malicious inclusions in hardware is becoming more importance due to the fact that most of our systems on chips are fabricated in Southeast Asia, although under the brand names of major U.S. companies. This can be explained by reducing production costs and increasing market competitiveness. Another side of a coin is losing a trust during a fabricating process. Especially, when it comes to development for military purposes, which may result in decommissioning weapon systems.

A model of compromised system is represented in Fig. 1.

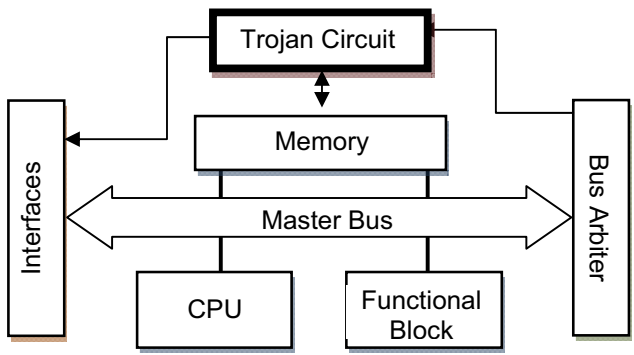

Fig. 1. Trojan insertion embedded in system on a chip

The trojan can be activated by a special value on Master Bus, for instance, it can be memory address where stored targeted data. Once trojan circuit is triggered, the payload can be one of the following: disabling system, transmitting interested data to third party by means of embedded interfaces, collecting accessed information in the memory for further utilization, rising security privileges for a current process running in the system.

### B. A Formal Model of Hardware Trojan (HT)

Let us consider a formal model of HT by introducing several abstract concepts. *Trojan ($T_i$)* is a malicious component that can provide an access to *System ($S_i$)* in certain moment with the appropriate condition.

The pairs $(T_i, O_i)$ are bound by the set of specified actions $A_s$. This set is defined according to security policy and specification of the vendor and is a subset of the whole set $A$ of all possible actions for each pair.

At the same time, pairs $(T_i, O_i)$ can communicate by a set of malicious actions $A_m$. It is obvious that $A = A_s \bigcup A_m$.

The purpose of security verification is identifying actions from the set $A_m$.

The task of malicious circuit detection is getting more complicated when HT can take advantage of a set of specified actions, that can gain an access to a computer system or its component, from the set $A_s$, such as $A_s \bigcap A_m \neq \varnothing$. As a result, it is needed to verify system considering a whole set of actions $A$. It is a hard verification task even for small systems on a chip because of searching within a set of all possible input vectors.

### C. Hardware Trojan Detection Task

The danger hidden in complex system on a chip nowadays is underestimated. The trojan circuit can be easily embedded to a system on a chip and hardly detected taking in consideration the size of the modern digital system [18].

The formal view to the problem of malicious insertions proves that the task of trojan detection in complex digital system is difficult.

The solution can be found in the area of high level testing methodology in order to cope with the complexity of the task. Nowadays there are powerful methods that are provided by researchers that can help in trojan detection and analysis, such as in [19] and [20], but still there is no mature solution that can provide universal methodology for fables companies and governments.

## IV. ASSESSING THE LOSSES OF THE COMPANY FROM SECURITY THREATS

The breaches in corporate environment may cause undesirable data leakage and will lead to suspending business processes of the company. In such scenario it may lose important customers and business partners because company which cannot protect itself from this attacks is faithless over the unforeseen costs like malicious programs influence, information drain, attacks on computer networks, etc [16].

The result is that when the number of personal computers is growing and communication channels capacity is increasing malware epidemic's scope and losses are growing correspondingly. Therefore the company management has to think about the information security.

In modern world the probability of malicious programs get into a computer system is constantly growing. It may cause not only short-term fault in the network, but a complete stopping the company. Losses by malicious programs are estimated as billions of dollars around the world annually and continue to increase.

According to [17] the cost of the average caused by malware attack in a corporate network can be calculated as in (1).

$$DELAY = \left(comp\_num \times fix\_time \times adjuster\_hour\_payment\right)$$
$$+ additional\_expenses +$$
$$+ \left(\frac{items\_day \times product\_price \times fix\_time \times comp\_num}{8 \times adjuster\_num}\right) + \quad (1)$$
$$+ \left(\frac{salary \times comp\_num \times fix\_time}{8 \times 22 \times adjuster\_num}\right),$$

where *comp_num* – number of computers within a network; *fix_time* – time in hours for fixing a fault; *adjuster_hour_payment* – payment for adjusting a computer per hour; *adjuster_num* – number of such specialists; *additional_expenses* – additional expenses for network repairing and buying new devices; *product_price* – price of a product; *items_day* – number of product items per day; *salary* – salary of an employee per month.

### V. ANALYSIS OF CORPORATE ANTIVIRUSES

According to latest report from Av-Comparatives Lab [1] the main players at corporate security market are Avira, Eset, G Data, Kaspersky, Sophos, and Symantec. In this article we will overview functional diversity of existed corporate suits and take a look to nearest future of corporate antivirus suits which seem to become a total security solution for corporate users.

The typical structure of corporate suite:
1) Administration console – provides useful managing and configuration environment for administrators of big networks.
2) Antivirus for workstation – actually the antivirus engine with all features peculiar to workstation antivirus. Provides centralized protection of user's system on a corporate network against all types of malware, network attacks, spam.
3) Mail server antivirus – protects the mail server against spam and malware delivered by email channels.
4) File server antivirus –protects data on servers under Microsoft Windows operation system control against all types of malware. Designed mostly for high-performance corporate servers.

Analyzing all products options it has been distinguished the main features of modern corporate antivirus:
1) Easy Installation and Deployment – simple and fast way to deploy the solution into a big corporate network, supporting Active Directory technology.
2) Usability and Management – console provides useful management interface with real-time monitoring and logging features.
3) Scalability – solution works with networks of different size from small business to enterprise scale with thousands of computers distributed geographically around many offices.
4) Technical Support and Updates – regularly delivers antivirus updates and helps to solve all unforeseen

security issues of a company with a short response time. Also website and online services are important points.
5) Cross-Platform Security – ability to protect systems with different types of operational systems, such as Linux, MacOS, mobile platforms, etc.

### VI. NOWADAYS TRENDS AND SUGGESTIONS

As for future in area of corporate users' security protection a growing trend is including more sophisticated administration interface that provides detailed information about the real-time status of the network. It can be represented as advanced graphical interface with diagrams or even as a separate product. It can be an intelligent agent that can handle huge amount of information from thousands of computers and hints the administrator what to do in that case.

For instance, Blue Medora designed a special agent for Symantec corporate solution which results in "less complexity, more uniform operations management, and a significant reduction in costs due to the elimination of redundant infrastructure and multiple platform-specific tools" [2]. It proves the idea that there is an area for further improvement of corporate antivirus solution even for the outstanding vendor.

Among extensible features are the following:
1) Improved monitoring of incidents with malware.
2) Improved monitoring of the user's intrusion into the antivirus key processes.
3) Monitoring of failures in updates and malware scanning tasks.

The new features to be included into the product:
1) Real-time status and availability monitor.
2) Log monitors.
3) Report and take-action system that would help administrator to perform necessary actions to any type of threats.

The main idea is to raise a sensitivity level of the persons who are responsible for corporate network security and reduce the time of reaction to the emerging danger. Therefore, useful and exhausted data representation can really help in struggling against malware.

Except logging and monitoring an essential part of security solution is integrity. Modern corporate antivirus solutions comprise not only a bunch - Antivirus, Antispyware, Firewall, Antispam with Managing System, but many additional features, such as Backup systems, Password and Key Managers and Encryption Utilities to organize safe confidential data storage.

This trend is peculiar to home solutions as well. Thus, Kaspersky Pure for home users provides besides malware protection also password management system to keep in safe all family's identities [3]. Another example, Norton Online Family also allows observing the kids activity on computer [4].

As for enterprise suits, Symantec provides Protection Suite for Endpoints where gathered encryption, confidential data storage and others features aimed to maintain IT security in a company [5]. One more interesting product came from Sophos [6]. Endpoint Security and Data Protection has Integrated DLP (Data Loss Prevention) and Encryption tools in its package.

Also mobile and non-Windows platforms should be supported within a corporate solution because of huge diversity of working devices: laptops, PDAs, smart phones, etc. Many antivirus vendors have such solutions in a product line.

The important point to be considered is Security-as-a-Service. A security is not only software, but a state of a system. It is important to have 24/7 technical support service to solve a newest security issues, such as new versions of malware, zero-day exploits. Often proactive defense cannot cope with a huge variety of new malware modification released every day by hacker's generators. The same way administrator cannot keep all software up-to-date with new patches installed. In such context deploying vulnerability searching system is desirable to reveal software breaches and notify to install new updates in time.

Here the problem of support service's quality has been raised. It is not a secret that a high quality support service can be granted only by the team of qualified malware experts not by "sandbox" robots [here we can put a reference to our research in "Sandbox Comparatives"]. Many companies provide malware analysis column on their web sites or even separate security domains where the descriptions of most popular threats are published, like it is done at virusradar.com by Eset and securelist.com by Kaspersky Lab.

Another side of the coin is an ability to remove consequences of an infection. Not all antivirus engines allow proper disinfection of the system or network after an incident that already has taken a place. In that case special removal utilities and scripts are released by analysts to help administrators in cleaning their IT farms. There are such services from Symantec [7] and AVZ tool from Kaspersky Lab [8].

Phishing is becoming a serious problem for all users in the cyber world. What antivirus vendors can suggest in protecting corporate users against this problem except of standard anti-phishing modules that block dangerous web sites from black list? The interesting solutions have been introduced within Kaspersky Internet Security 2011 – Geo Filter and Online Banking modules. According to information from official site: "Geo Filter provides the user with an option to block domains related to specific countries. Online Banking controls requests to Online Banking services while processing confidential data" [9]. Those modules could be helpful in keeping a communication with financial institutions more safe which could be essential in corporate environment.

Finally, a following to corporate security standards is what some AV vendors do. The big companies try to organize corporate IT security according to policies compliant to security standards. Among them:

1)  X509 – is ITU-T standard specifies formats for public key certificates, certificate revocation lists, attribute certificates, and a certification path validation algorithm [10],

2)  LDAP (Lightweight Directory Access Protocol) – is an application protocol for querying and modifying data using directory services running over TCP/IP [11],

3)  Microsoft IWA (Integrated Windows Authentication) – provides authentication connections between Microsoft IIS, Internet Explorer, and other Active Directory aware applications [12].

## VII. CONCLUSION

To sum up, in this short review the current security threats have been briefly presented. According to them an analysis of antivirus solution for corporate users was proposed. The general features and structure of corporate suit were enumerated based on the latest report from av-comparatives.org. In the last part of the article we considered modern trends in current antivirus solutions from most popular AV vendors, such as Eset, Symantec, Sophos and Kaspersky.

It is obvious that the corporate products represent quite powerful solutions for enterprise networks but they could become better by adopting new standards, technologies and a high level of support services. The corporate suit is becoming a heavy package of tools aimed to fight against malware, network attacks, spam, phishing. It gives to administrators a control under a huge corporate network that allows monitoring a real-time activity and react to an existing situation as soon as possible. A corporate security is a multifactor system that consists of security software, services, policies and a human factor. None of them should be missed in a building process of secure corporate environment.

## REFERENCES

[1]  "Review of IT Security Suites for Corporate Users", May 2009. Available: www.av-comparatives.org.

[2]  Blue Medore Agent for Symantec Endpoint Protection, Available: http://www.bluemedora.com/product/page/40

[3]  "Kaspersky Pure. Ultimate Protection for Your Digital Life", Available: http://www.kaspersky.com/kaspersky-pure

[4]  "Norton™ Online Family. A smarter way to keep your kids safe online". Available: https://onlinefamily.norton.com/familysafety/loginStart.fs

[5]  "Symantec Protection Suite Enterprise Edition for Endpoints". Available: http://www.symantec.com/business/protection-suite-enterprise-edition-for-endpoints

[6]  Sophos, "Endpoint Security and Data Protection". Available: http://www.sophos.com/products/enterprise/endpoint/security-and-control/

[7]  Symantec, "Spyware and Virus Removal". Available: http://www.symantec.com/norton/nortonlive/spyware-virus-removal.jsp

[8]    "How to scan your computer, save the log and run a script using the
       AVZ utility? ". Available:
       http://support.kaspersky.com/faq/?qid=208279710
[9]    Kaspersky Internet Security 2011 Manual. Available:
       http://support.kaspersky.com/kis2011?level=2
[10]   Wikipedia, http://en.wikipedia.org/wiki/X.509
[11]   Wikipedia, http://en.wikipedia.org/wiki/LDAP
[12]   Wikipedia,
       http://en.wikipedia.org/wiki/Integrated_Windows_Authentication
[13]   Microsoft Security Intelligence Report. Volume8. July-Dec2009,
       Available: http://www.microsoft.com/security/about/sir.aspx
[14]   Malware Encyclopedia. Available:
       http://www.totalmalwareinfo.com/eng/Net-
       Worm.Win32.Kido_/_Conficker.A-C_Worm
[15]   "PC giant warns of hardware trojan", NewScientist, 22 July 2010.
[16]   Filatov, Kozlovskih, Cvetkova, Planning, Finance, Management of
       Enterprise. – Finance and Statistics, 2005, 384 p.
[17]   M.V. Bocharinkova, A. S. Saprykin, V.A. Kiktenko, A. S. Adamov,
       "Developing methods to assess damage from the spread of malware
       at enterprises", IT-Security Conference for New Generation,
       Moscow, Russia, 28-29 April 2009.
[18]   X. Wang, M. Tehranipoor and J. Plusquellic, "Detecting Malicious
       Inclusions in Secure Hardware: Challenges and Solutions",
       International Workshop on Hardware Oriented Security and Trust,
       2008, pp. 15-22.
[19]   I. Verbauwhede, P. Schaumont, "Design methods for Security and
       Trust", DATE'07, 2007.
[20]   F. Wolff, C. Papachristou, S. Bhunia, R. S. Chakraborty, "Towards
       Trojan-Free Trusted ICs: Problem Analysis and Detection Scheme",
       DATE'08, 2008.